%% file: main_IV.tex
\documentclass[letterpaper, 10 pt, conference]{ieeeconf}
\IEEEoverridecommandlockouts                          

\usepackage{array}
\usepackage{textcomp}
\usepackage{stfloats}
\usepackage{verbatim}
\usepackage{cite}
\usepackage{xcolor}

\usepackage[colorlinks=true,linkcolor=black,citecolor=black,urlcolor=blue]{hyperref}

\usepackage{microtype}
\usepackage{graphicx}
\usepackage{subfigure}
\usepackage{booktabs} 
\usepackage{bbm}

\usepackage{pgfplots}

\usepackage{amsmath,amssymb,amsfonts}
\usepackage{xurl}
\usepackage{stackengine}
\usepackage{tikz}
\usetikzlibrary{decorations.pathreplacing}
\usetikzlibrary{positioning,arrows.meta,quotes}
\usetikzlibrary{shapes,snakes}
\usetikzlibrary{bayesnet}
\tikzset{>=latex}
\tikzstyle{plate caption} = [caption, node distance=0, inner sep=0pt, below left=5pt and 0pt of #1.south]

\usepackage[normalem]{ulem}
\usepackage{multirow}

\begin{document}
\title{Learning Car-Following Behaviors Using Bayesian \\Matrix Normal Mixture Regression}

\author{Chengyuan~Zhang$^1$,~\IEEEmembership{Graduate Student Member,~IEEE,}%
~Kehua~Chen$^{2,3}$,~\IEEEmembership{Graduate Student Member,~IEEE,}\\%
~Meixin~Zhu$^{4,5}$,~\IEEEmembership{Member,~IEEE,}%
~Hai~Yang$^{3}$,%
~and~Lijun~Sun$^1$,~\IEEEmembership{Senior Member,~IEEE,}
\thanks{(Corresponding author: Meixin Zhu)}
\thanks{$^\dagger$ Chengyuan Zhang and Kehua Chen contributed equally to this work.}
\thanks{$^1$C. Zhang and L. Sun are with the Department of Civil Engineering, McGill University, Montreal, QC H3A 0C3, Canada. E-mail addresses:
enzozcy@gmail.com (C. Zhang), lijun.sun@mcgill.ca (L.
Sun).}
\thanks{$^2$K. Chen is with Division of Emerging Interdisciplinary Areas (EMIA), Academy of Interdisciplinary Studies, The Hong Kong University of Science and Technology, Hong Kong, China.}%
\thanks{$^3$K. Chen and H. Yang are also with Department of Civil and Environmental Engineering, The Hong Kong University of Science and Technology, Hong Kong, China.}%
\thanks{$^4$M. Zhu is with Intelligent Transportation Thrust, Systems Hub, The Hong Kong University of Science and Technology (Guangzhou), Guangzhou, China.}%
\thanks{$^5$M. Zhu is also with Guangdong Provincial Key Lab of Integrated Communication, Sensing and Computation for Ubiquitous Internet of Things, Guangzhou, China.}
}


\maketitle

\begin{abstract}
Learning and understanding car-following (CF) behaviors are crucial for microscopic traffic simulation. Traditional CF models, though simple, often lack generalization capabilities, while many data-driven methods, despite their robustness, operate as ``black boxes" with limited interpretability. To bridge this gap, this work introduces a Bayesian Matrix Normal Mixture Regression (MNMR) model that simultaneously captures feature correlations and temporal dynamics inherent in CF behaviors. This approach is distinguished by its separate learning of row and column covariance matrices within the model framework, offering an insightful perspective into the human driver decision-making processes. Through extensive experiments, we assess the model's performance across various historical steps of inputs, predictive steps of outputs, and model complexities. The results consistently demonstrate our model's adeptness in effectively capturing the intricate correlations and temporal dynamics present during CF. A focused case study further illustrates the model's outperforming interpretability of identifying distinct operational conditions through the learned mean and covariance matrices. This not only underlines our model's effectiveness in understanding complex human driving behaviors in CF scenarios but also highlights its potential as a tool for enhancing the interpretability of CF behaviors in traffic simulations and autonomous driving systems.
\end{abstract}


\IEEEpeerreviewmaketitle

\section{Introduction}
Car-following (CF) maneuver involves the interactive response of the following vehicles (FV) dictated by the states of the leading vehicles (LV) in the longitudinal direction \cite{zhang2023interactive}. 
As a vital component in microscopic traffic analysis, accurately modeling CF behaviors benefits traffic management and the development of autonomous driving. Therefore, researchers have dedicated efforts to CF models over the past several decades \cite{saifuzzaman2014incorporating, treiber2013microscopic, zhu2018modeling}.

Early studies developed models based on traffic theory and human behaviors, such as Gipps' model \cite{gipps1981behavioural}, Newell model \cite{newell2002simplified}, and Intelligent Driver Model (IDM) \cite{treiber2000congested}. Although traditional models have clearly defined formulas with high interpretability, these models have low complexity and cannot capture complex temporal dynamics. With the thriving of data-driven methods, a number of studies utilized advanced deep learning techniques to model CF behaviors, such as deep reinforcement learning \cite{zhu2018human} and Recurrent Neural Networks (RNNs) \cite{wang2017capturing}. While the aforementioned data-driven approaches attain satisfactory accuracy, it is important to note that they function as black-box models with limited interpretability, and lack the capability to quantify uncertainties.

To capture the stochasticity of human behaviors while maintaining reasonable interpretability, some studies modeled CF behaviors from Bayesian perspectives, such as using the Gaussian Mixture Model (GMM) \cite{angkititrakul2009evaluation, lefevre2014comparison, angkititrakul2013stochastic}. GMM is widely adopted since it explicitly models the uncertainty in the parameters, providing a better understanding of the model's confidence in its predictions. This is crucial in applications with safety-critical driving tasks, where understanding the uncertainty of the model can help human drivers make better decisions. Besides, apart from GMM capturing stochasticity, it is also good at depicting nonlinear human behaviors with explainable modalities. For instance, Angkititrakul et al. \cite{angkititrakul2009evaluation} employed GMM and piecewise auto-regressive exogenous algorithms to model CF behaviors. 
Afterward, the authors enhance the model structure, enabling non-parametric Bayesian modeling of CF behaviors \cite{angkititrakul2013stochastic}.
Lefevre et al. \cite{lefevre2014comparison} conducted a comparison between parametric and non-parametric methods for CF modeling. They observed that the relative performance of the various models exhibited significant variations depending on the prediction horizon. Recently, Chen et al. \cite{chen2023bayesian} used a Bayesian GMM to achieve probabilistic modeling of CF. In their study, they developed an inference algorithm using Gibbs sampling and provided an analysis of model interpretability. Based on GMM, Zhang et al. \cite{zhang2023interactive} proposed a framework to quantify the interactions between the FV and LV.

Nonetheless, vanilla GMM learns the full covariance matrices given historical information, while there often exist explicit correlations among the variables or in the temporal space. This provides us an insight that a separable covariance structure of the proposed model can effectively reduce the model parameters, speed up computations, and improve practical interpretability. With these motivations, we develop a probabilistic CF model with good interpretability using Bayesian Matrix Normal Mixture Regression (MNMR). Specifically, we first learn a joint distribution in terms of the Bayesian Matrix Normal Mixture Model (MNMM) \cite{viroli2011finite, yin2023analysis}, then make predictions based on conditional distributions. The learned row and column covariance matrices indicate the correlations among features and temporal dynamics, respectively. Compared with the previous data-driven models, our proposed MNMR provides an intuitive understanding of complex CF dynamics. We conduct extensive experiments using the HighD dataset and present a case study to illustrate the interpretability of our model. 

To sum up, the contributions of this work are as follows:
\begin{itemize}
    \item We present a novel model named Bayesian Matrix Normal Mixture Regression (MNMR) to authentically capture car-following dynamics while appropriately accounting for stochasticity.
    \item We perform extensive experiments using the HighD dataset to demonstrate and analyze the impact of different historical steps, prediction steps, and model complexity in the regression task.
    \item We present a case study to demonstrate the outperforming interpretability of our model, showcasing that the learned mean and covariance matrices of different components represent distinct human driving behaviors.
\end{itemize}

\section{Preliminaries and Problem Definition}\label{problem_def}
The primary objective of this work is to develop a comprehensive statistical model based on Bayesian learning to accurately represent, interpret, and analyze CF behaviors. This model aims to capture the complex, dynamic interactions between a follower-leader pair in a traffic stream, with a particular focus on the following aspects: (1) The variability in car-following patterns under different traffic situations; and (2) The temporal dynamics of car-following, considering factors like speed, distance, and relative acceleration. 

In a typical CF task, an FV adjusts its driving speed based on the actions of the leading vehicle (LV). Given a continuous period (e.g., the past $T$ time steps) of observations, this work aims to predict the FV's actions in the next $\Delta T$ time steps. Here we denote $s_{(\cdot)}(t)$, $v_{(\cdot)}(t)$, and $a_{(\cdot)}(t)$ as the position, speed, and acceleration of vehicles at a specific time $t$, where $(\cdot)$ can be either the FV or LV. In the setting of a regression task, we consider the following regressors for CF modeling at each time step $t$: the speed of FV (i.e., $v_{\mathrm{FV}}(t)$), the speed difference between FV and LV (i.e., $\Delta v =v_\mathrm{FV}(t)-v_\mathrm{LV}(t)$), and the longitudinal spacing between FV and LV (i.e., $\Delta s = s_\mathrm{FV}(t)-s_\mathrm{LV}(t)-\mathrm{vehicle\,length}$), denoted as $\boldsymbol{x}(t)\in \mathbb{R}^{D_x=3}$. We select the action of FV (i.e., $a_\mathrm{FV}(t)$) as the response variable $y(t)\in \mathbb{R}^{D_y=1}$. In general, the CF regression task is to learn a mapping $f: \boldsymbol{x}(t)\in\mathbb{R}^{D_x} \mapsto \boldsymbol{y}(t)\in\mathbb{R}^{D_y}$, but note that in our setting $D_y=1$, indicating that $y$ is a scalar.

In literature, Gaussian mixture regression (GMR) provides an elegant method to estimate the distribution of output data given input data \cite{wang2018learning}. The elemental assumption of GMR is that all the data points are generated from a GMM with unknown parameters. Therefore, it encodes the joint distribution of input and output using a GMM
\begin{align}
    \left[\begin{array}{c}
         \boldsymbol{x}(t)\\
         \boldsymbol{y}(t)  
    \end{array}\right]\sim\sum_{k=1}^K\pi_k\mathcal{N}(\boldsymbol{\mu}^{(t)}_k,\boldsymbol{\Sigma}^{(t)}_k),
\end{align}
where $\boldsymbol{\mu}^{(t)}_k\in\mathbb{R}^{(D_x+D_y)}$ and $\boldsymbol{\Sigma}^{(t)}_k\in\mathbb{R}^{(D_x+D_y)\times(D_x+D_y)}$. Then it exploits the Gaussian conditioning theorem to build the mapping from input to output (i.e., $p(\boldsymbol{y}(t)|\boldsymbol{x}(t))$). In practice, although it is efficient to implement GMR at each time step, the need to estimate the full covariance matrices $\boldsymbol{\Sigma}^{(t)}_k$ would become complicated when dealing with time series. For instance, once we have observed $\boldsymbol{x}\in \mathbb{R}^{(D_x\times T)}$ from the past $T$ time steps, we then want to predict the future driving behaviors $\boldsymbol{y}'\in \mathbb{R}^{(D_y\times \Delta T)}$. Correspondingly, the parameters of GMM are adapted to $\boldsymbol{\mu}^{(t)}_k\in\mathbb{R}^{(D_x\times T+D_y\times \Delta T)}$ and $\boldsymbol{\Sigma}^{(t)}_k\in\mathbb{R}^{(D_x\times T+D_y\times \Delta T)\times(D_x\times T+D_y\times \Delta T)}$. As aforementioned, learning the full covariance matrices requires more computational resources with longer time series.

To relieve the computational burdens from learning the full covariance matrices, in this work, we model the observations as a matrix variable instead of a vector-form and leverage the trick of the Kronecker product to decompose a full covariance matrix into two smaller covariance matrices. 

\section{Method}\label{method}
Accurately model and analyze human driving data is inherently complex due to the high dimensionality and intricate temporal dynamics. Traditional models often struggle to capture the full scope of the correlations and nonlinear dynamics. In this section, we introduce a Bayesian Matrix Normal Mixture Regression (MNMM) and a Matrix Normal Mixture Regression (MNMR) to effectively model such data.

\subsection{Matrix Normal Mixture Model (MNMM)}
First, as aforementioned, we model the observations as a matrix-variable instead of a vector one. Here, we assume that spatiotemporal driving data $\boldsymbol{\mathcal{X}}\in\mathbb{R}^{N\times D\times \tau}$ can be modeled by the matrix normal distribution. For the notation conciseness, we have denoted $D=D_x+D_y$ and $\tau = T+\Delta T$. Let the matrix-variate symbol $\boldsymbol{X}$ be a $D \times \tau$ random matrix following a matrix normal distribution with mean matrix $\boldsymbol{M}\in\mathbb{R}^{D\times \tau}$ to represent the data trends, row covariance matrix $\boldsymbol{U}\in\mathbb{R}^{D\times D}$ to capture the correlations among variables, and column covariance matrix $\boldsymbol{V}\in\mathbb{R}^{\tau\times \tau}$ to depict the temporal dynamics, whose dual structure allows for a more explicit understanding of the driving behavior trends. Then we have,
\begin{align}
    \boldsymbol{X} \sim \, & \, \mathcal{MN}_{}(\boldsymbol{M},\boldsymbol{U},\boldsymbol{V}),\\
    \Rightarrow \mathrm{vec}(\boldsymbol{X}) \sim \, & \, \mathcal{N}(\mathrm{vec}(\boldsymbol{M}), \boldsymbol{V}\otimes\boldsymbol{U}),
\end{align}
where $\mathrm{vec}(\cdot)$ is the vectorization operator that stacks all the columns in a matrix into a tall column vector. To distinguish the matrix normal distribution and multivariate normal distribution, we represent them using $\mathcal{MN}$ and $\mathcal{N}$, respectively. In addition, the symbol $\otimes$ denotes the Kronecker product, defined as
\begin{align}
    \boldsymbol{V} \otimes \boldsymbol{U} =\begin{bmatrix}v_{11}\boldsymbol{U} &\cdots &v_{1\tau}\boldsymbol{U} \\\vdots &\ddots &\vdots \\v_{\tau1}\boldsymbol{U} &\cdots &v_{\tau\tau}\boldsymbol{U} \end{bmatrix}\in\mathbb{R}^{(D\times \tau)\times(D\times \tau)}.
\end{align}
Note that $\forall \xi\neq0$, a property that Kronecker product holds is $\boldsymbol{V}\otimes\boldsymbol{U} = (\boldsymbol{V}/\xi)\otimes(\xi\boldsymbol{U})$. Therefore, to avoid the non-identifiable inference issue, we put a strong prior on one covariance matrix and an uninformative prior on another.


Here we define a mixture model of matrix normal distributions with $K$ components \cite{viroli2011finite}:
\begin{align}
    \boldsymbol{X} \, \sim \, \sum_{k=1}^K \pi_k \mathcal{MN}(\boldsymbol{M}_k,\boldsymbol{U}_k,\boldsymbol{V}_k),
\end{align}
where $0\leq\pi_k\leq1$ denote the mixing weight of the $k$-th matrix normal distribution, and $\sum_{k=1}^K \pi_k=1$.

\begin{figure}
    \centering
    {\centering
    \resizebox{0.80\linewidth}{!}{\input{Figs/MNMM_pgm}}}
    \caption{Probabilistic graphical model of MNMM.}\label{pgm}
    \vspace{-0.5cm}
\end{figure}
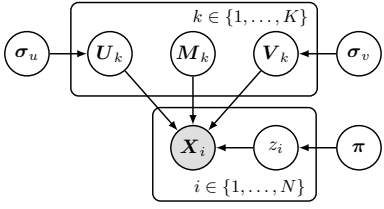

In a Bayesian framework, we develop a MNMM as follows
\begin{align}
    &\mathbf{For\,\,component}\,k = 1,\ldots,K:\nonumber\\
    &\quad \quad \boldsymbol{M}_k \sim \mathcal{MN}(\boldsymbol{M}_0,\boldsymbol{U}_0,\boldsymbol{V}_0),\\
    &\quad \quad \boldsymbol{V}_k \sim \mathrm{LKJCholeskyCov}(\eta, \boldsymbol{\sigma}_v),\\
    &\quad \quad \boldsymbol{U}_k \sim \mathrm{LKJCholeskyCov}(\eta, \boldsymbol{\sigma}_u),\\
    &\mathbf{For\,\,sample}\,i = 1,\ldots,N:\nonumber\\
    &\quad \quad \boldsymbol{z}_i \sim \mathrm{Cat}(\boldsymbol{\pi}),\label{Eq_z_cat}\\
    &\quad \quad \boldsymbol{X}_i \sim \, \mathcal{MN}(\boldsymbol{M}_{z_i},\boldsymbol{U}_{z_i},\boldsymbol{V}_{z_i}),\label{Eq_X_MN}
\end{align}
where `LKJ' represents the LKJ distribution \cite{lewandowski2009generating}, and the priors are defined as $\boldsymbol{\pi} \sim \mathrm{Dirichlet}(\boldsymbol{\alpha})$, $\boldsymbol{\sigma}_v \sim \mathrm{Exp}(\lambda_v)$ and $\boldsymbol{\sigma}_u \sim \mathrm{Exp}(\lambda_u)$, in which $\boldsymbol{\alpha}$, $\boldsymbol{M}_0$, $\boldsymbol{U}_0$, $\boldsymbol{V}_0$, $\eta$, $\lambda_v$, $\lambda_u$, are all hyperparameters to tune. The corresponding model structure is illustrated as a probabilistic graphical model, see Fig.~\ref{pgm}.

\subsection{Matrix Normal Mixture Regression (MNMR)}
In an interactive driving scenario, human drivers make decisions and perform driving maneuvers based on observed information from several past seconds \cite{wang2022social}. This insight reveals that it is significant to incorporate the past observations into the future predictions. Therefore, based on this insight, we develop a Matrix Normal Mixture Regression (MNMR) based on MNMM in the following.

In a sequence-to-sequence case, we are interested in learning a mapping $f: \boldsymbol{\tilde{X}} = [\boldsymbol{X}, \boldsymbol{X}']\in\mathbb{R}^{D_x\times (T+\Delta T)} \mapsto \boldsymbol{\tilde{Y}} = [\boldsymbol{Y},\boldsymbol{Y}']\in\mathbb{R}^{D_y\times (T+\Delta T)}$, where $\boldsymbol{X} \in\mathbb{R}^{D_x\times T}$ , $\boldsymbol{X}' \in\mathbb{R}^{D_x\times \Delta T}$, $\boldsymbol{Y} \in\mathbb{R}^{D_y\times T}$, and $\boldsymbol{Y}' \in\mathbb{R}^{D_y\times \Delta T}$. Then the MNMM can be rewritten as:
\begin{align}
    \left[\begin{array}{c}
         \boldsymbol{\tilde{X}}\\
         \boldsymbol{\tilde{Y}}  
    \end{array}\right] = \left[\begin{array}{cc}
         \boldsymbol{X}& \boldsymbol{X}'\\ 
         \boldsymbol{Y}&\boldsymbol{Y}' 
    \end{array}\right] \sim \, \sum_{k=1}^K \pi_k \mathcal{MN}\left(\boldsymbol{M}_k, \boldsymbol{U}_k, \boldsymbol{V}_k\right),
\end{align}
where
\begin{align}
    \boldsymbol{M}_k&=\left[\begin{array}{c}
         \boldsymbol{M}_{\tilde{X},\,k} \\
         \boldsymbol{M}_{\tilde{Y},\,k} 
    \end{array}\right]=\left[\begin{array}{cc}
         \boldsymbol{M}_{X,\,k} & \boldsymbol{M}_{X',\,k}\\ 
         \boldsymbol{M}_{Y,\,k} & \boldsymbol{M}_{Y',\,k} 
    \end{array}\right],\nonumber\\
    \boldsymbol{V}_k&=\left[\begin{array}{cc}
         \boldsymbol{V}_{11,\,k}\in\mathbb{R}^{T\times T}&\boldsymbol{V}_{12,\,k}\in\mathbb{R}^{T\times \Delta T} \\ 
         \boldsymbol{V}_{21,\,k}\in\mathbb{R}^{\Delta T\times T}&\boldsymbol{V}_{22,\,k}\in\mathbb{R}^{\Delta T\times \Delta T}
    \end{array}\right],\nonumber\\
    \boldsymbol{U}_k&=\left[\begin{array}{cc}
         \boldsymbol{U}_{11,\,k}\in\mathbb{R}^{D_x\times D_x}&\boldsymbol{U}_{12,\,k}\in\mathbb{R}^{D_x\times D_y} \\ 
         \boldsymbol{U}_{21,\,k}\in\mathbb{R}^{D_y\times D_x}&\boldsymbol{U}_{22,\,k}\in\mathbb{R}^{D_y\times D_y}
    \end{array}\right].\nonumber
\end{align}

This form is equivalent to the vectorized form
\begin{align}
     \left[\begin{array}{c}
         \mathrm{vec}(\boldsymbol{\tilde{X}}) \\
         \mathrm{vec}(\boldsymbol{\tilde{Y}})
    \end{array}\right] \sim \, \sum_{k=1}^K \pi_k \mathcal{N}\left(\left[\begin{array}{c}
         \mathrm{vec}(\boldsymbol{M}_{\tilde{X},\,k}) \\ \mathrm{vec}(\boldsymbol{M}_{\tilde{Y},\,k}) 
\end{array}\right], \boldsymbol{V}_k\otimes\boldsymbol{U}_k \right),
\end{align}
and for $\boldsymbol{V}_k\otimes\boldsymbol{U}_k$ we have
\begin{align}
    \boldsymbol{V}_k\otimes\boldsymbol{U}_k = \left[\begin{array}{cc}
         \boldsymbol{V}_{11,\,k}\otimes\boldsymbol{U}_k & \boldsymbol{V}_{12,\,k}\otimes\boldsymbol{U}_k \\ 
         \boldsymbol{V}_{21,\,k}\otimes\boldsymbol{U}_k & \boldsymbol{V}_{22,\,k}\otimes\boldsymbol{U}_k
    \end{array}\right].
\end{align}

The target of the matrix normal mixture regression (MNMR) model is deriving $p(\boldsymbol{\tilde{Y}}|\boldsymbol{\tilde{X}})$, therefore, the joint distribution can be decomposed into two parts
\begin{align}
    \mathrm{vec}(\boldsymbol{\tilde{X}}) \sim &\, \sum_{k=1}^K \pi_k \mathcal{N}\left(\mathrm{vec}(\boldsymbol{M}_{\tilde{X},\,k}), \boldsymbol{V}_{11,\,k} \otimes \boldsymbol{U}_k \right),\\
    \mathrm{vec}(\boldsymbol{\tilde{Y}}) | \mathrm{vec}(\boldsymbol{\tilde{X}}) \sim &\, \sum_{k=1}^K w_k \mathcal{N}\left( \mathrm{vec}(\boldsymbol{M}_{\tilde{Y}|\tilde{X},\,k}), \boldsymbol{\Sigma}_{\tilde{Y}|\tilde{X},\,k}\right),\label{conditional}
\end{align}
where we have:
\begin{footnotesize}  
\begin{align}
    &w_k = \frac{\pi_k \mathcal{N}\left(\mathrm{vec}(\boldsymbol{\tilde{X}})|\mathrm{vec}(\boldsymbol{M}_{\tilde{X},\,k}), \boldsymbol{V}_k\otimes \boldsymbol{U}_{11,\,k} \right)}{\sum_{k=1}^K \pi_k \mathcal{N}\left(\mathrm{vec}(\boldsymbol{\tilde{X}})|\mathrm{vec}(\boldsymbol{M}_{\tilde{X},\,k}), \boldsymbol{V}_k\otimes \boldsymbol{U}_{11,\,k} \right)},\\
   & \mathrm{vec}(\boldsymbol{M}_{\tilde{Y}|\tilde{X},\,k}) = \mathrm{vec}(\boldsymbol{M}_{\tilde{Y},\,k}) + \nonumber\\ &\qquad\qquad(\boldsymbol{V}_{21,\,k}\otimes\boldsymbol{U}_k)(\boldsymbol{V}_{11,\,k}^{-1}\otimes\boldsymbol{U}_k^{-1})(\mathrm{vec}(\boldsymbol{\tilde{X}})-\mathrm{vec}(\boldsymbol{M}_{\tilde{X},\,k})),\\
    &\boldsymbol{\Sigma}_{\tilde{Y}|\tilde{X},\,k} =  \boldsymbol{V}_{22,\,k}\otimes\boldsymbol{U}_k - \nonumber\\&\qquad\qquad(\boldsymbol{V}_{21,\,k}\otimes\boldsymbol{U}_k)(\boldsymbol{V}_{11,\,k}^{-1}\otimes\boldsymbol{U}_k^{-1})(\boldsymbol{V}_{12,\,k}\otimes\boldsymbol{U}_k).
\end{align}
\end{footnotesize}

Similar to the above procedures, it is easy to derive $p(\mathrm{vec}(\boldsymbol{Y}') | \mathrm{vec}(\boldsymbol{Y}), \mathrm{vec}(\boldsymbol{\tilde{X}}))$ based on $p(\mathrm{vec}(\boldsymbol{\tilde{Y}}) | \mathrm{vec}(\boldsymbol{\tilde{X}}))$. Furthermore, given that we cannot observe the future information $\boldsymbol{X}'$ at a specific time step, and we still want to predict the future $\boldsymbol{Y}'$ given the previous observations, we can therefore marginalize $\boldsymbol{X}'$ to obtain 
\begin{align}\nonumber
    &p(\mathrm{vec}(\boldsymbol{Y}') | \mathrm{vec}(\boldsymbol{Y}), \mathrm{vec}(\boldsymbol{X}))\\
    =&\int_{\boldsymbol{X}'}p(\mathrm{vec}(\boldsymbol{Y}') | \mathrm{vec}(\boldsymbol{Y}), \mathrm{vec}(\boldsymbol{\tilde{X}}))d\boldsymbol{X}',
\end{align}
which is still a Gaussian mixture model
\begin{align}
    \mathrm{vec}(\boldsymbol{Y}') | \mathrm{vec}(\boldsymbol{Y}), \mathrm{vec}(\boldsymbol{X})\sim \sum_{k=1}^K \beta_k\mathcal{N}(\boldsymbol{\mu}_{y,k}, \boldsymbol{\Sigma}_{y,k}),\label{conditioned_gmm}
\end{align}
where the mean vector $\boldsymbol{\mu}_{y,k}\in\mathbb{R}^{(D_y\times \Delta T)}$ and the covariance matrix $\boldsymbol{\Sigma}_{y,k}\in\mathbb{R}^{(D_y\times \Delta T)\times (D_y\times \Delta T)}$.

\section{Experiments}\label{experiments}

\begin{table*}[t]\centering
\caption{Regression performance on the acceleration observations.}
\scriptsize
\begin{tabular}{c|c|c|c|c|c|c|c|c|c|c|c|c|c|c|c|c}\toprule
\multicolumn{2}{c}{\multirow{2}{*}{}} &\multicolumn{5}{|c|}{RMSE on the test data / $K$}& \multicolumn{5}{c|}{MAE on the test data / $K$} & \multicolumn{5}{c}{NLL on the training data / $K$} \\
\cmidrule{1-17}
$\Delta T$ & $T$ &$5$ &$10$ &$20$ &$40$ &$60$ &$5$ &$10$ &$20$ &$40$ &$60$ &$5$ &$10$ &$20$ &$40$ &$60$ \\
\midrule
\multirow{5}{*}{1} & 1 & 0.054 & 0.055 & 0.054 & 0.053 & 0.053 &
 0.047 & 0.047 & 0.046 & 0.046 & 0.045 &
 -0.213 & -0.240 & -0.234 & -0.247 & -0.271 \\
 &3 & 0.013 & 0.014 & 0.013 & 0.013 & 0.013 &
 0.011 & 0.011 & 0.011 & 0.011 & 0.011 &
 -1.611 & -1.603 & -1.665 & -1.677 & -1.716 \\
 &5 & 0.019 & \textbf{0.012} & 0.013 & \textbf{0.012} & \textbf{0.011} &
 0.015 & \textbf{0.010} & 0.011 & \textbf{0.010} & \textbf{0.010} &
 -1.757 & -1.877 & -2.043 & -2.116 & -2.084 \\
 &7 & 0.015 & 0.014 & 0.020 & 0.015 & 0.016 &
 0.013 & 0.012 & 0.017 & 0.012 & 0.013 &
 -2.006 & -2.138 & -2.237 & -2.255 & -2.344 \\
 &9 & 0.025 & 0.016 & 0.018 & 0.017 & 0.013 &
 0.021 & 0.014 & 0.015 & 0.014 & 0.011 &
 -2.191 & -2.248 & -2.327 & \textbf{-2.440} & -2.333 \\
\cmidrule{1-17}
\multirow{5}{*}{3} & 1 & 0.240 & 0.230 & 0.223 & 0.219 & 0.205 &
 0.107 & 0.104 & 0.101 & 0.099 & 0.093 &
 -1.547 & -1.611 & -1.648 & -1.674 & -1.696 \\
&3 & 0.115 & 0.089 & 0.106 & \textbf{0.069} & \textbf{0.069} &
 0.045 & 0.037 & 0.043 & \textbf{0.030} & \textbf{0.030} &
 -1.994 & -2.007 & -2.026 & -2.188 & -2.172 \\
&5 & 0.250 & 0.110 & 0.107 & 0.122 & \textbf{0.080} &
 0.097 & 0.045 & 0.044 & 0.048 & \textbf{0.033} &
 -1.861 & -2.054 & -2.157 & -2.256 & -2.297 \\
&7 & 0.189 & 0.129 & 0.126 & 0.095 & 0.140 &
 0.077 & 0.053 & 0.053 & 0.039 & 0.055 &
 -2.113 & -2.326 & -2.333 & \textbf{-2.432} & -2.376 \\
 &9 & 0.174 & 0.148 & 0.167 & 0.149 & 0.120 &
 0.077 & 0.060 & 0.071 & 0.063 & 0.050 &
 -2.219 & -2.302 & -2.290 & -2.383 & -2.360 \\
\cmidrule{1-17}
\multirow{5}{*}{5} & 1 & 0.772 & 0.788 & 0.759 & 0.652 & 0.675 &
 0.249 & 0.237 & 0.250 & 0.214 & 0.217 &
 -1.987 & -1.831 & -2.083 & -2.088 & -2.114 \\
 & 3 & \textbf{0.220} & 0.321 & 0.376 & \textbf{0.222} & 0.380 &
 \textbf{0.068} & 0.093 & 0.107 & \textbf{0.068} & 0.110 &
 -2.399 & -2.169 & -2.181 & -2.339 & -2.271 \\
 &5& 0.361 & 0.536 & 0.323 & \textbf{0.241} & 0.262 &
 0.103 & 0.130 & 0.095 & \textbf{0.072} & 0.077 &
 -2.081 & -2.209 & -2.100 & \textbf{-2.429} & \textbf{-2.408} \\
 &7& 0.395 & 0.248 & \textbf{0.235} & 0.331 & 0.317 &
 0.123 & 0.076 & 0.072 & 0.096 & 0.096 &
 -2.088 & -2.226 & -2.316 & -2.377 & -2.381 \\
 &9 & 0.334 & 0.391 & 0.349 & 0.321 & 0.281 &
 0.097 & 0.117 & 0.102 & 0.093 & 0.083 &
 -2.314 & -2.228 & -2.353 & \textbf{-2.444} & \textbf{-2.412} \\
\cmidrule{1-17}
 \multirow{5}{*}{7} & 1 & 1.612 & 1.429 & 1.018 & 1.093 & 0.934 &
 0.452 & 0.379 & 0.283 & 0.309 & 0.262 &
 -2.162 & -2.082 & -2.140 & -2.187 & -2.334 \\
 & 3 & 0.632 & 0.856 & 0.818 & 0.490 & 0.685 &
 0.155 & 0.222 & 0.210 & 0.123 & 0.161 &
 -2.177 & -2.311 & -2.381 & -2.393 & -2.418 \\
 &5& \textbf{0.433} & 0.708 & 0.559 & 0.500 & \textbf{0.451} &
 \textbf{0.113} & 0.159 & 0.135 & 0.129 & \textbf{0.113} &
 -2.185 & -2.330 & -2.380 & -2.344 & -2.380 \\
&7& 0.591 & 0.521 & \textbf{0.483} & \textbf{0.483} & 0.525 &
 0.152 & 0.134 & \textbf{0.118} & \textbf{0.123} & 0.127 &
 -2.328 & -2.391 & -2.246 & \textbf{-2.447} & \textbf{-2.474} \\
 &9 & 0.549 & 0.590 & 0.615 & 0.503 & \textbf{0.478} &
 0.142 & 0.143 & 0.154 & 0.130 & \textbf{0.121} &
 -2.373 & -2.277 & \textbf{-2.427} & \textbf{-2.492} & \textbf{-2.467} \\
\cmidrule{1-17}
 \multirow{5}{*}{9} & 1 & 1.194 & 1.227 & 1.210 & 1.281 & 1.233 &
 0.292 & 0.299 & 0.302 & 0.306 & 0.306 &
 -2.151 & -2.307 & -2.253 & -2.409 & -2.446 \\
 & 3 & 0.910 & 0.875 & 1.143 & 1.044 & 0.849 &
 0.208 & 0.206 & 0.251 & 0.231 & 0.177 &
 -2.254 & -2.259 & -2.293 & -2.300 & -2.290 \\
 & 5 & 0.950 & \textbf{0.665} & \textbf{0.683} & 0.998 & \textbf{0.716} &
 0.208 & \textbf{0.143} & \textbf{0.155} & 0.216 & \textbf{0.158} &
 -2.185 & -2.310 & -2.487 & -2.445 & \textbf{-2.512} \\
 & 7 & 0.761 & 0.957 & 0.853 & \textbf{0.716} & \textbf{0.742} &
 0.171 & 0.200 & 0.180 & \textbf{0.158} & \textbf{0.163} &
 -2.456 & -2.434 & -2.475 & -2.489 & \textbf{-2.523} \\
 & 9 & 0.850 & \textbf{0.731} & 0.830 & 1.132 & 0.811 &
 0.221 & \textbf{0.156} & 0.178 & 0.240 & 0.175 &
 -2.356 & \textbf{-2.515} & \textbf{-2.504} & \textbf{-2.621} & \textbf{-2.577} \\
\bottomrule
\multicolumn{8}{l}{{}}
\label{Tab:res}
\end{tabular}
\vspace{-0.5cm}
\end{table*}

\subsection{Dataset and Preprocessing}
Experiments are performed on the HighD dataset that contains high-resolution naturalistic vehicle trajectories extracted from drone videos of German highways. Compared to the commonly used NGSIM dataset \cite{punzo2011assessment}, the HighD dataset benefits from more reliable data capture methods and advanced computer vision techniques. It features 60 recordings captured at different times of the day, ranging from 8:00 to 17:00, and has a resolution of 25 Hz. In our experiment, the original dataset is downsampled to a smaller set with a sampling frequency of $5$ Hz, achieved by uniformly selecting every $5$-th sample. HighD dataset provides detailed information on the trajectories, velocities, and accelerations of vehicles, which is essential for developing and evaluating car-following models that accurately capture real-world driving scenarios. In this study, we follow the same data processing procedures as in \cite{zhang2021spatiotemporal} to transform the data into a new coordinate system. We selected 123 leader-follower pairs where the car-following duration lasted for more than 50 sec. Of these pairs, $93$ pairs ($\sim75\%$) were used to train our model, while the remaining $30$ pairs ($\sim25\%$) were stored as the test set to evaluate the forecasting performance. By using pairs with longer car-following durations, we aim to capture more realistic driving behaviors and enable our model to better handle complex and dynamic traffic situations.

\subsection{Regression Performance Evaluation}
Based on the proposed Bayesian structure, we employ Stochastic Variational Inference (SVI) based on Pyro to infer the posterior distributions \cite{bingham2018pyro}. The model is trained using the Adam optimizer with a learning rate of 0.01. The experiments were conducted on a Linux server with one NVIDIA RTX A4000 GPU.

In general, a trade-off in selecting the number of components $K$ arises: with too many components, the learned model may over-fit the data, while a model with too few components may not be flexible enough to represent an underlying distribution of observations. To evaluate the performance impact of factors from different aspects, we conduct extensive experiments using combinations of the predicted steps $\Delta T$, the past steps $T$, and the number of components $K$. Specifically, we set $\Delta T=\{1, 3, 5, 7, 9\}$ steps, $T=\{1, 3, 5, 7, 9\}$ steps, and $K=\{5, 10, 20, 40, 60\}$. For each combination of $\{T,\Delta T, K\}$, the regression performance is evaluated using the root mean square error (RMSE) and the mean absolute error (MAE) on the test set, and the negative log-likelihood (NLL) on the training data. These metrics are recognized as effective in diagnosing the variation in the errors in a set of forecasts \cite{willmott1985statistics}. 

Table \ref{Tab:res} presents the evaluation results, in which we mark the relatively good performances with bold text. Note that in this table, the lower the value, the better the results. Generally, the prediction errors decrease with longer historical steps while increasing with longer prediction steps. As for the model complexity, the errors would decrease initially but followed by an increase with the increase of $K$, which is consistent with the aforementioned patterns: A small value of $K$ is inadequate for capturing the complex dynamics of CF, while a large $K$ tends to overfit the training samples, thereby reducing the model's generalization ability and has worse performance when evaluating on the test data. This point is especially obvious when dealing with longer prediction steps and large $K$ --- the performance on the training set seem perfect but bad on the test set.

\begin{figure}[t]
    \centering
    \subfigure{
        \centering \includegraphics[width=\linewidth]{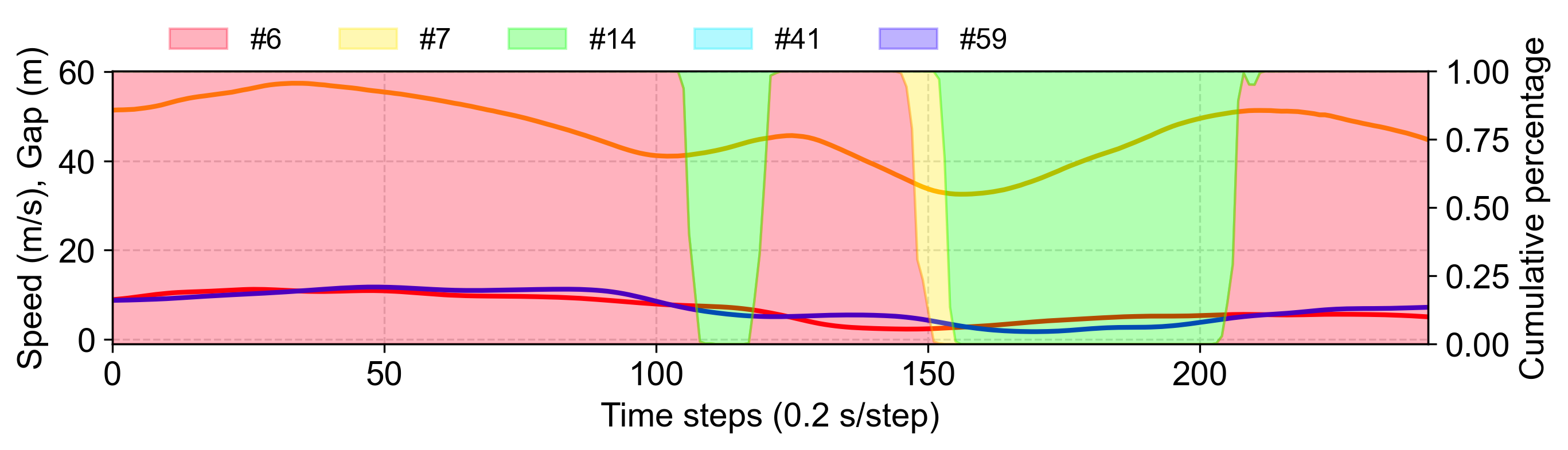}
    }
    \subfigure{
        \centering \includegraphics[width=\linewidth]{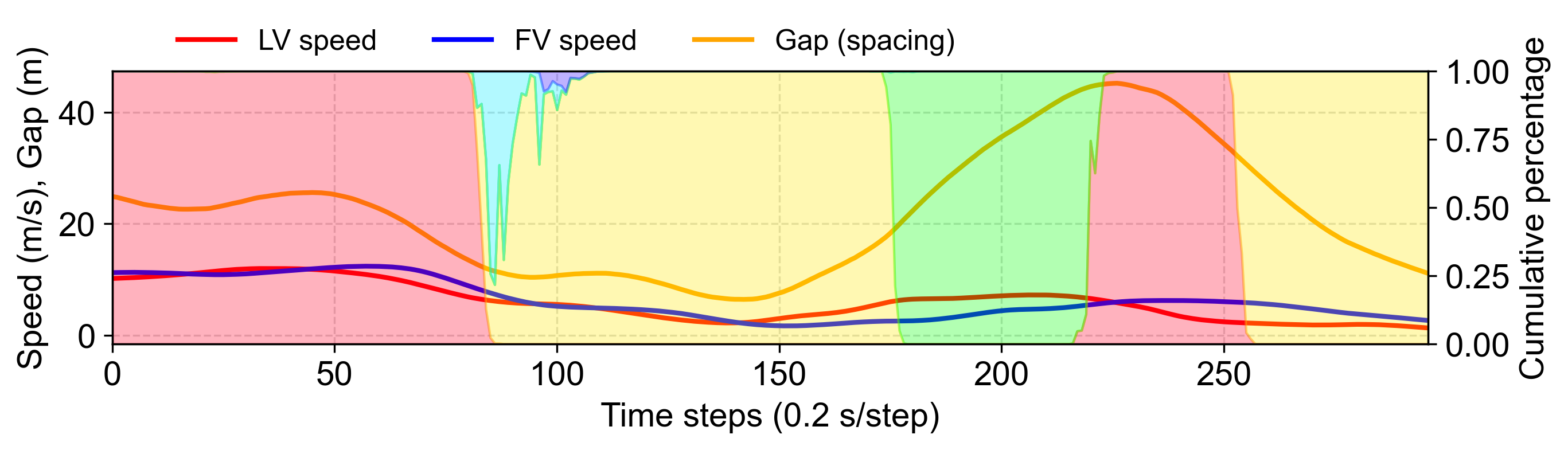}
    }
    \vspace{-0.5cm}
    \caption{Two illustrative cases to understand the dominant components (i.e., $\#6$, $\#7$, $\#14$, $\#41$, and $\#59$ in these cases) and the transitioning among them.}
    \label{fig:component}
    
\end{figure}

\begin{figure}[t]
    \centering
    \includegraphics[width=\linewidth]{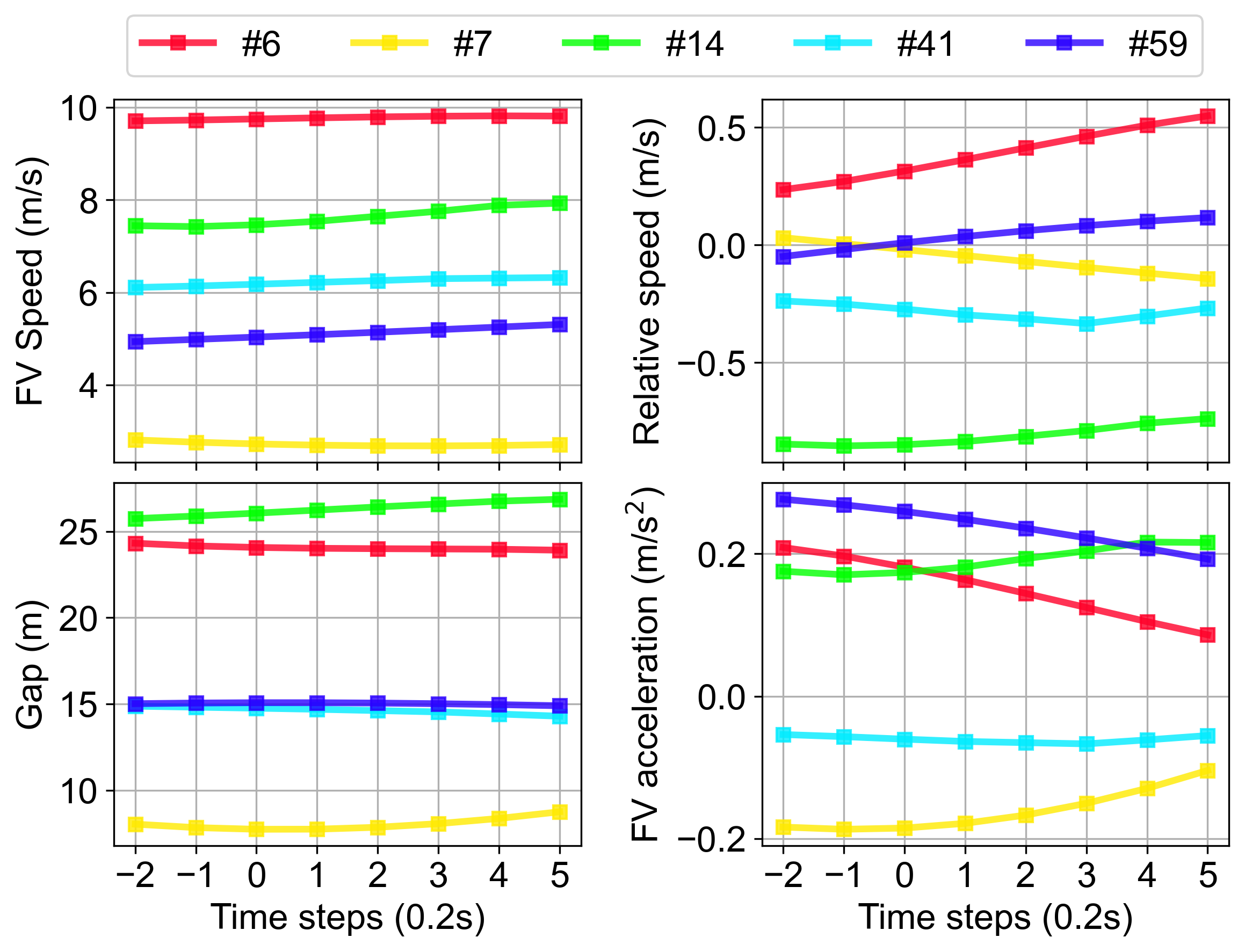}
    \vspace{-0.5cm}
    \caption{The mean matrix $\boldsymbol{M}_k$ of the dominant components.}
    \label{fig:mean}
    \vspace{-0.5cm}
\end{figure}

\begin{figure}[t]
    \centering
    \subfigure[$\#6$]{
        \centering \includegraphics[width=.48\textwidth]{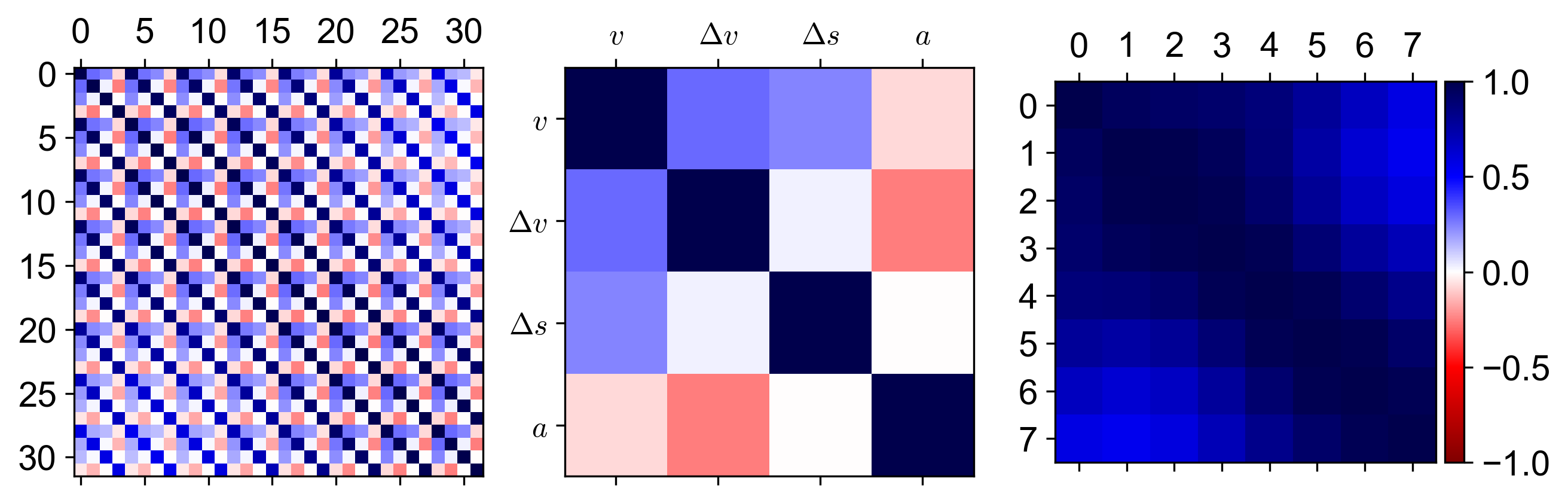}
    }\\
    \subfigure[$\#7$]{
        \centering \includegraphics[width=.48\textwidth]{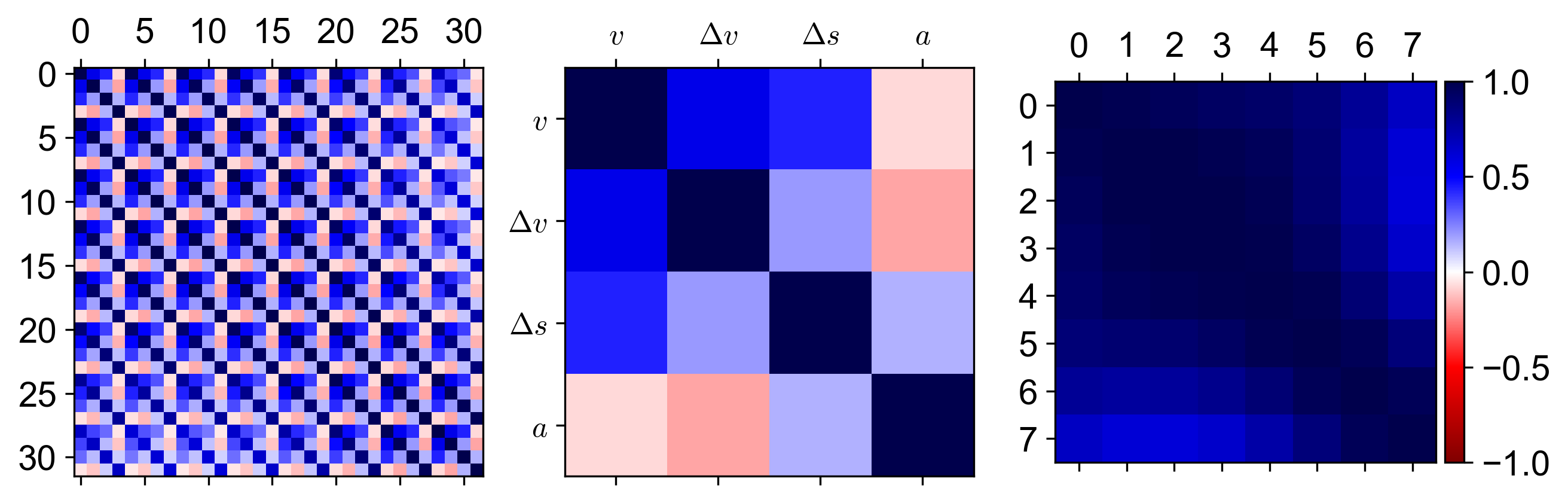}
    }\\
    \subfigure[$\#14$]{
        \centering \includegraphics[width=.48\textwidth]{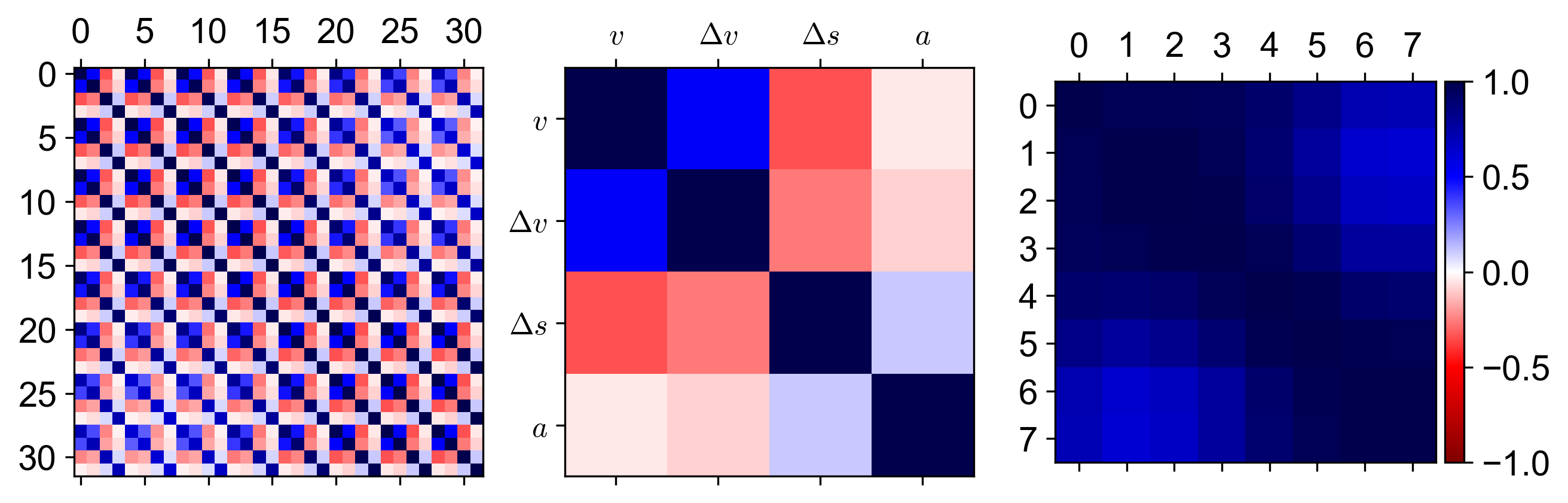}
    }\\
    \subfigure[$\#41$]{
        \centering \includegraphics[width=.48\textwidth]{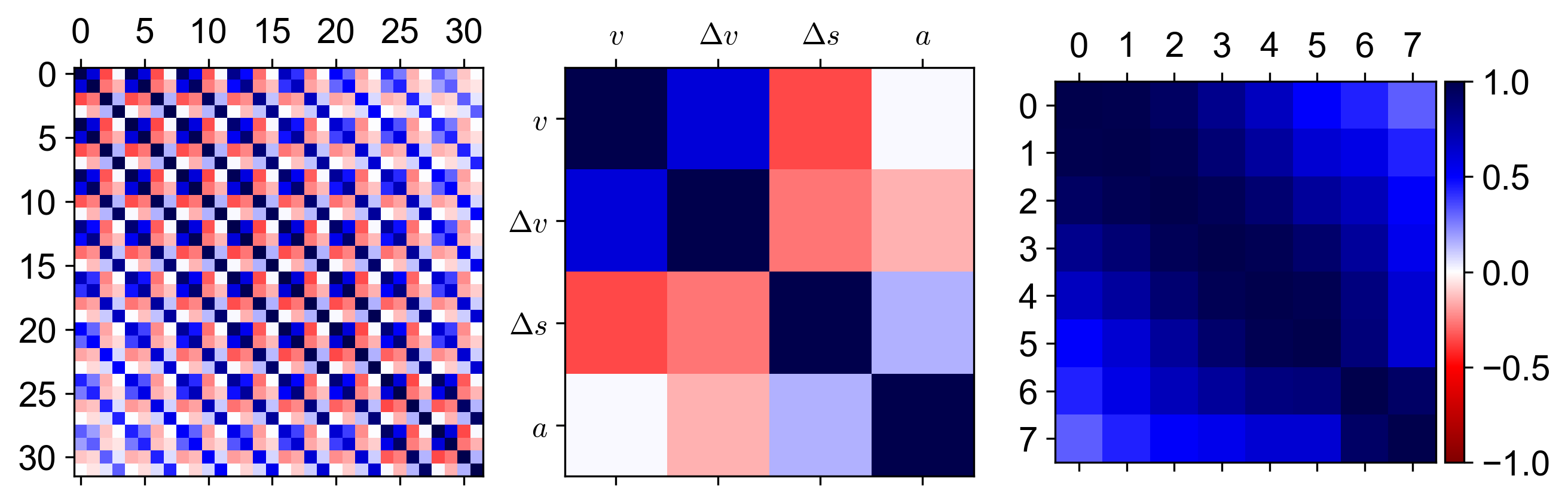}
    }\\
    \subfigure[$\#59$]{
        \centering \includegraphics[width=.48\textwidth]{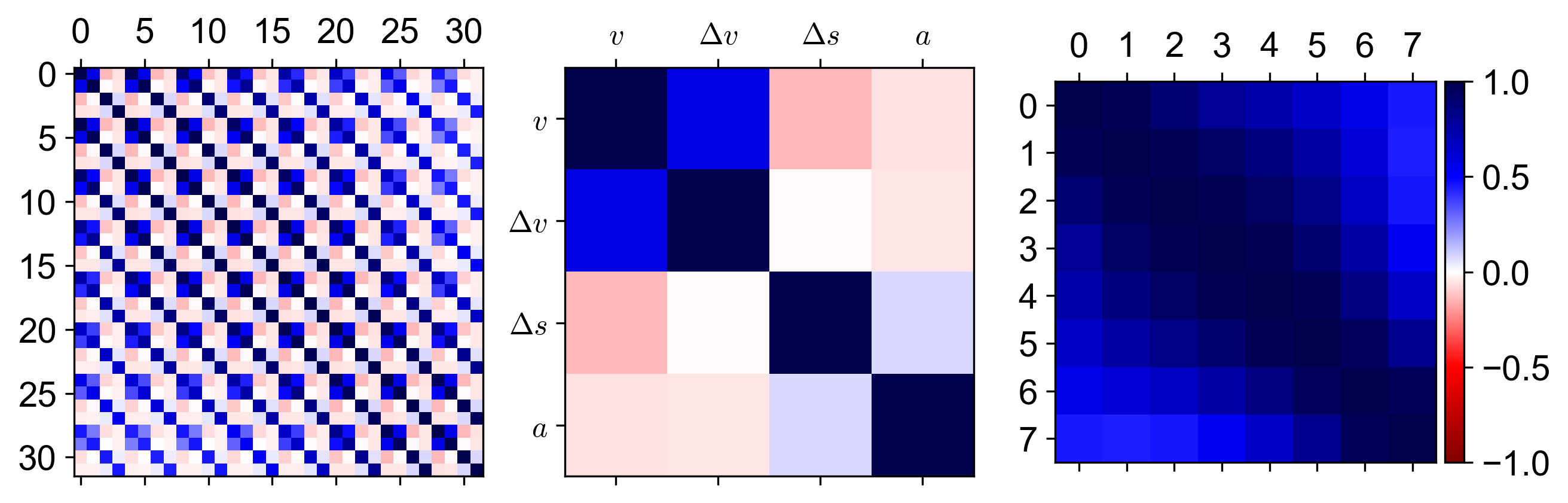}
    }\\
    \caption{The correlation matrices of the dominant components.}
    \label{fig:matrix}
    \vspace{-0.5cm}
\end{figure}

\subsection{Mixture Components Interpreting}
Here we highlight that MNMR, as a mixture model, can provide an intuitive understanding of CF behaviors. Based on the model performance and following the settings in \cite{chen2023bayesian}, we choose a case with $\{T=5, \Delta T=3, K=60\}$ as a light demonstration. Fig. \ref{fig:component} shows two cases with the variation of five dominant components, in which the LV speed, FV speed, and spacing gap are shown in solid lines; also, the percentage of $\beta_k$ in Eq.~\eqref{conditioned_gmm} are shown as the shaded area. Fig. \ref{fig:mean} illustrates the mean matrix $\boldsymbol{M}_k$ of dominant components.
We can see that component $\#6$ exerts dominance in the beginning 110 steps and 80 steps for Case 1 (the upper part) and Case 2 (the lower part), respectively. In these instances, FVs exhibit large speeds, gaps, and relative speeds, which are also shown in the trends in the mean matrix illustrated in Fig. \ref{fig:mean}. During these stages, the FV slightly accelerates to keep a closer distance to the LV, which leads to the increasing relative speed. Besides, component $\#14$ emerges as a significant component during time steps 110-120 and 160-200 in Case 1. During these periods, the FV's action indicates a typical accelerating stage --- the LV rapidly increases its speed (especially when from a congestion stage to a free-flow stage) and the FV also accelerates to follow up the FV. But due to the action delay, although the FV accelerates hard, it still has a negative relative speed, i.e., $\Delta v =v_\mathrm{FV}(t)-v_\mathrm{LV}(t)<0$, to the LV. 
Regarding component $\#7$, it dominates during time steps 90-180 and 260-300 in Case 2. Such scenarios reveal the slow-moving stage when the FV and LV are close to each other at a very low speed, and the FV decelerates to keep a safe distance from the LV.

Fig. \ref{fig:matrix} visualize the learned correlation matrices of the dominant components. Recall that in MNMR, we independently learn the row and column covariance matrices. Here, the left column denotes the full correlation matrix of $\boldsymbol{V}\otimes\boldsymbol{U}$, and the plots in the second and third columns depict the feature and temporal correlation matrices, respectively.
For components $\#6$ and $\#7$, we observe that the speed and relative speed are negatively correlated with acceleration, aligning with our common sense that when speeds and relative speeds are large, FVs are more likely to decelerate. Additionally, gaps are positively correlated to both speed and relative speed since large gaps provide room for FVs to speed up.
Conversely, components $\#14$, $\#41$, and $\#59$ suggest that speed and relative speed are negatively correlated with both gaps and acceleration. This observation shows that these two components correspond to operational conditions where FVs are approaching LVs, and FVs tend to decelerate to maintain a safe distance. Concerning the temporal correlation matrix, it is noteworthy that considering neighboring time steps is significant in modeling human driving behaviors, given that the correlations consistently remain positive. Nevertheless, the correlations diminish as the temporal distances increase --- according to the results, the most correlated elements are those located within 1 second before and after the diagonal, which is consistent with the results discovered before \cite{zhang2024bayesian}.

\section{Conclusion}\label{conclusion}
In the realm of understanding complex human decision-making processes, particularly in car-following (CF) scenarios, the role of generative models is indispensable. These models, e.g., GMM, adept at capturing the joint probability distribution of various inputs and outputs, offer a window into the intricate interplay of causes and effects that define human behavior. This paper introduces a novel approach to modeling human driving behaviors in CF situations through the development of a Bayesian Matrix Normal Mixture Model (MNMM). Our model stands out by employing a Kronecker structure within the covariance matrix, a technique that effectively decouples feature correlations from temporal dynamics. This unique attribute allows for a nuanced representation of the interdependencies inherent in the driving process.

Building on the foundational principles of MNMM, we extend our methodology to encompass Matrix Normal Mixture Regression (MNMR). This progression enables the prediction of future driving actions based on past observations, mirroring the decision-making patterns of human drivers who rely on recently acquired information. Our comprehensive evaluation, conducted using the HighD dataset, demonstrates the model's exceptional capability to encapsulate complex driving dynamics. A case study is presented to further illustrate the superior interpretability of our approach.

Our work has broad implications for advancing planning and decision-making algorithms of autonomous driving by modeling and understanding human behaviors, offering insights to enhance safety and efficiency. It not only contributes a novel theoretical model but also paves the way for practical applications in the field of automotive engineering. 





\bibliographystyle{IEEEtran}
\bibliography{IEEEabrv, sample}
%



\end{document}

%% file: Figs/MNMM_pgm.tex
\usetikzlibrary{positioning,arrows.meta,quotes}
\usetikzlibrary{shapes,snakes}
\usetikzlibrary{bayesnet}
\tikzset{>=latex}
\tikzstyle{plate caption} = [caption, node distance=0, inner sep=0pt,
below left=5pt and 0pt of #1.south]
\tikzset{every picture/.style={line width=0.75pt}} 

\begin{tikzpicture}

    \node [circle,draw=black,inner sep=0pt,minimum size=0.8cm] (Sigma) at (-1.5-10,1) {$\boldsymbol{U}_k$};

    \node [circle,draw=black,inner sep=0pt,minimum size=0.8cm] (V) at (+1.5-10,1) {$\boldsymbol{V}_k$};
    
    \node [circle,draw=black,inner sep=0pt,minimum size=0.8cm] (mu) at (0-10,1) {$\boldsymbol{M}_k$};
    
    \node [circle,draw=black,inner sep=0pt,minimum size=0.8cm] (pik) at (+3-10,-.7) {$\boldsymbol{\pi}$};
    
    \node [circle,draw=black,inner sep=0pt,minimum size=0.8cm] (zi) at (+1.5-10,-.7) {$z_i$};
	
    \node [obs,draw=black,inner sep=0pt,minimum size=0.8cm] (xi) at (-0-10,-.7) {$\boldsymbol{X}_{i}$};

    
    \plate [color=black,inner sep=.3cm,xshift=0.02cm,yshift=0.2cm] {part3} {(V)(mu)(Sigma)} {};
    
    \node [circle,draw=black,inner sep=0pt,minimum size=0.8cm] (hyper_Sigma) at (-3-10,1) {$\boldsymbol{\sigma}_u$};

    \node [circle,draw=black,inner sep=0pt,minimum size=0.8cm] (hyper_Sigma_v) at (+3-10,1) {$\boldsymbol{\sigma}_v$};
    
    \path [draw,->] (hyper_Sigma) edge (Sigma);
    \path [draw,->] (hyper_Sigma_v) edge (V);
    
	
    \path [draw,->] (mu) edge (xi);
    \path [draw,->] (Sigma) edge (xi);
    \path [draw,->] (V) edge (xi);
    
    \path [draw,->] (zi) edge (xi);
    
    \plate [color=black,inner sep=.3cm,xshift=0.0cm,yshift=0cm] {part1} {(zi)(xi)} {};
    
    \node [text width=2.5cm] (m) at (1.27-10,-1.42) {\small{$i\in\{1,\dots,N\}$}};
    
    \node [text width=5cm] (k1) at (+2.5-10,1.66) {\small{$k\in\{1,\dots,K\}$}};
    
    \path [draw,->] (pik) edge (zi);
    
    
\end{tikzpicture}